\documentclass[aps,pra,preprint,superscriptaddress]{revtex4-1}

\usepackage{amsmath}
\usepackage{amssymb}
\usepackage{mathenv}
\usepackage[]{subfigure}
\usepackage{epsfig}

\begin{document}


\title{Theoretical and experimental analysis of rare earth whispering gallery mode laser relative intensity noise}


\author{Jean-Baptiste Ceppe}
\affiliation{FOTON, CNRS UMR 6082, Universit\'{e} de Rennes 1, ENSSAT, 6 rue de Kerampont, 22300 Lannion, France}
\author{Michel Mortier}
\affiliation{PSL Research University, Chimie ParisTech - CNRS, Institut de Recherche de Chimie de Paris, Paris 75005, France}
\author{Patrice F\'{e}ron}
\author{Yannick Dumeige}
\email[]{yannick.dumeige@univ-rennes1.fr}
\affiliation{FOTON, CNRS UMR 6082, Universit\'{e} de Rennes 1, ENSSAT, 6 rue de Kerampont, 22300 Lannion, France}

\date{\today}

\begin{abstract}
\noindent The relative intensity noise (RIN) of a solid state whispering-gallery-mode class-B laser is studied both theoretically and experimentally under different pumping regimes. In particular, we show that harmonics of the spiking frequency are observed in the RIN spectrum. A rate equation model including Langevin forces and the nonlinear coupling between inverted ion and photon number fluctuations has been developed to reproduce the experimental results and to extract relevant physical parameters from the fitting of the RIN spectrum.
\end{abstract}


\maketitle

\section{Introduction}

\noindent Solid state microcavity lasers have a lot of applications in integrated photonics or microwave photonics \cite{Ilchenko06, Chiasera10}. Among available configurations, rare earth doped whispering-gallery-mode (WGM) micro-resonators can be used as narrow linewidth or low threshold miniaturized from visible to mid infrared  lasers \cite{Sandoghdar96, Lissillour01a, He13, Deng14, Palma17} due to their low volume and high quality (Q) factors. High purity microwave sources can be obtained using the beating of WGM lasers \cite{Xiao10}. Solid-state micro-lasers can also serve as sensitive biological or chemical integrated sensors \cite{He2010, Maker13, Bernhardi13, Palma16, Li17}. A lot of geometries (spheres, torus \cite{Yang03}, disks \cite{Kippenberg06}) and materials have been studied since the first demonstration of solid state WGM micro-resonator lasing \cite{Sandoghdar96}. These particular lasers have been extensively studied in terms of oscillation threshold \cite{Sandoghdar96, Maker13b}, output power performances \cite{Ristic16} and spectral purity \cite{Lissillour01}. Otherwise the phase or frequency noise of semiconductor or fiber lasers including a WGM resonator has been studied \cite{Collodo14, Liang15}. Nevertheless, to date only few papers report on the dynamical properties of WGM miniaturized lasers: the auto-modulation regime \cite{Schwartz07} of a toroidal WGM laser have been demonstrated \cite{He10} and an extensive theoretical of intensity and frequency noise of micro-resonator Brillouin lasers have been published \cite{Loh15}. However, detailed study of the relative intensity noise (RIN) of WGM micro-lasers have not been reported. The intensity noise performance knowledge is of great importance for communication applications based on amplitude modulation \cite{Tan13}. In the purpose of sensing applications based on the use of micro-lasers, the RIN of the source is a crucial parameter since it will determine the sensitivity of the whole device. Moreover, microcavity laser noise properties investigation is of interest from a fundamental point of view \cite{Bjork94, Kilper97, Takemura12, Lebreton13} nevertheless, to our knowledge, neither theoretical nor experimental small mode volume WGM laser RIN analysis have been carried out so far.

In this paper we report on the RIN measurements of Erbium doped fluoride glass WGM micro-lasers for several pumping configurations. Due to the unique properties of WGM resonators, harmonics of the spiking frequency are observed in the RIN spectrum. Usual models of class-A or -B vertical cavity surface emitting laser RIN \cite{Baili08, Baili09} have been extended to take into account WGM specificities in order to exploit the experimental results. The paper is organized as follows. We first present our model detailing its specificity. Then we describe the experimental setup and the studied fluoride glass WGM micro-resonators. Finally we discuss some comparisons between theoretical and experimental results in the last section.

\section{Model}

\noindent We first describe the model used in this paper to analyse the RIN experimental results. The model is based on the coupled rate equations driven by noise sources. We will show that for WGM micro-lasers the linearisation usually made to solve the set of coupled differential equations in the frequency domain can not be used. In this section we propose an iterative method similar to the perturbation theory \cite{Rosencher02} to obtain an approximated solution to the coupled nonlinear differential equation system. 

\subsection{Notations}

\noindent The Fourier transform of a signal $f(t)$ will be denoted $\hat{f}(\omega)=\mathrm{FT}\left[f(t)\right](\omega)$. To lighten the calculation developments and to make easier the application of the Wiener-Khintchine theorem \cite{Mandel95} we will use two different definitions of the operator $\left\langle \right\rangle$. In the time domain for a stationary signal $f(t)$ it is defined as 
\begin{equation}
\left\langle f(t)\right\rangle=\lim\limits_{T\rightarrow +\infty}\frac{1}{T}\int_{-T/2}^{T/2}f(t)dt.
\end{equation}
\noindent In the frequency domain, $\left\langle \right\rangle$ will be related to the Fourier transform of two signals $f(t)$ and $g(t)$ by
\begin{equation}
\left\langle \hat{f}(\omega)\hat{g}^*(\omega)\right\rangle=\lim\limits_{T\rightarrow +\infty}\frac{1}{T}\mathbb{E}\left[\hat{f}_T(\omega)\hat{g}_T^*(\omega)\right]
\end{equation}
where $\hat{g}^*(\omega)$ denotes the conjugate of $\hat{g}(\omega)$. $f_T(t)=f(t)$ if $t\in[-T/2,T/2]$ anf $f_T(t)=0$ otherwise. $\mathbb{E}$ represents here the mean expectation of a random variable.

\subsection{Laser modelling}

\noindent The Erbium ions (density $\mathcal{N}$) will be modelled by an atomic 3-level system with an excited lifetime $T_1=1/A$ (where $A$ is the average probability of spontaneous emission) and an emission cross section $\sigma_e$. We consider here a single-mode laser characterized by a mode volume $V_{m}$ and a photon lifetime $\tau_{ph}$. $n_0$ is the refractive index of the cavity material. $W_P$ represents the pumping rate. The dynamic evolution of a class-B laser is given by the following set of coupled differential equations where $\Delta N(t)$ is the inverted ion number and $F(t)$ the intracavity photon number \cite{Siegman86, De13}
\begin{EqSystem}
\frac{d \Delta N}{d t}=\frac{\Delta N_0-\Delta N}{\tau}-2\kappa F\Delta N\label{syst1a}\\ 
\frac{d F}{d t}=-\frac{F}{\tau_{ph}}+\kappa F\Delta N\label{syst1b}
\end{EqSystem}
where we define the recovery time $\tau=1/(W_P+A)$, the coupling between inverted ion and photon numbers
\begin{equation}\label{cappa}
\kappa=\frac{c\sigma_e}{n_0V_{m}},
\end{equation}
and the unsaturated inverted ion number
\begin{equation}
\Delta N_0=N\frac{W_P-A}{W_P+A},
\end{equation}
with $N$ representing the total ion numbers $N=\mathcal{N}V_m$.

\subsection{Static solution above threshold}

\noindent The first step of the calculation consists to solve Eqs (\ref{syst1a}) and (\ref{syst1b}) in the stationary regime by setting $\frac{d \Delta N}{d t}=0$ an $\frac{d F}{d t} =0$. Above threshold, the inversion population value $\overline{\Delta N}$ is clamped to the value
\begin{equation}
\overline{\Delta N}=\Delta N_{th}=\frac{1}{\kappa\tau_{ph}},
\end{equation}
where $\Delta N_{th}$ is the inversion population at threshold. Which gives the stationary value for the intracavity photon number
\begin{equation}
\overline{F}=\frac{\eta-1}{2\kappa\tau},
\end{equation}
where we have defined $\eta=\frac{\Delta N_0}{\Delta N_{th}}$. We can also evaluate the threshold pumping rate by
\begin{equation}
W_{P,th}=A\frac{N+\Delta N_{th}}{N-\Delta N_{th}},
\end{equation}
and define the pumping rate $r=\frac{W_P}{W_{P,th}}$.

\subsection{Small signal analysis}

\noindent In this section we calculate the fluctuations of the photon number around its stationary value $\overline{F}$. In this aim we define inversion population and photon number fluctuations $\delta\Delta N(t)$ and $\delta F(t)$ by
\begin{subequations}
\begin{eqnarray}
\Delta N(t) & = & \overline{\Delta N}+\delta\Delta N(t)\label{Fluc1}\\
F(t) & = & \overline{F}+\delta F(t)\label{Fluc2}.
\end{eqnarray}
\end{subequations}
The usual approach consists to inject Eqs (\ref{Fluc1}) and (\ref{Fluc2}) in Eqs (\ref{syst1a}) and (\ref{syst1b}) and to linearise the system. $\widehat{\delta \Delta N}(\omega)$and $\widehat{\delta F}(\omega)$ can be thus found algebraically by taking the Fourier transform of the whole linearized system. In the present case we want to take into account intracavity nonlinear effects and this method cannot be applied. Nevertheless the nonlinearities can be considered iteratively by applying a perturbation development of $\delta\Delta N(t)$ and $\delta F(t)$
\begin{subequations}
\begin{eqnarray}
\delta \Delta N(t) & = & \lambda\delta\Delta N^{(1)}(t)+\lambda^2\delta\Delta N^{(2)}(t)\label{pertu1}\\
\delta F(t) & =& \lambda \delta F^{(1)}(t)+\lambda^2 \delta F^{(2)}(t)\label{pertu2},
\end{eqnarray}
\end{subequations}
where we assume $\left\langle \delta \Delta N^{(1)}(t)\right\rangle=0$, $\left\langle \delta F^{(1)}(t)\right\rangle=0$, $\left\langle \delta \Delta N^{(2)}(t)\right\rangle=0$ and $\left\langle \delta F^{(2)}(t)\right\rangle=0$. These two equations are injected in Eqs (\ref{Fluc1}), (\ref{Fluc2}), (\ref{syst1a}) and (\ref{syst1b}), by identifying the first order terms in $\lambda$ we obtain the following set of differential equations
\begin{EqSystem}
\frac{d \delta \Delta N^{(1)}}{d t}=-\frac{\eta}{\tau}\delta \Delta N^{(1)}-\frac{2}{\tau_{ph}}\delta F^{(1)}+\xi_N(t)\label{Per1a}\\
\frac{d \delta F^{(1)}}{d t}=\frac{\eta-1}{2\tau}\delta \Delta N^{(1)}+\xi_F(t)\label{Per1b},
\end{EqSystem}
where we have added driving Langevin forces $\xi_N(t)$ and $\xi_F(t)$ to take into account all of excess noise sources. We assume that these processes represent two white Gaussian noises with $\left\langle \xi_N(t)\right\rangle=0$ and $\left\langle \xi_F(t)\right\rangle=0$ which are characterized by their autocorrelation functions
\begin{subequations}
\begin{eqnarray}
\left\langle\xi_N(t)\xi_N^*(t+\tau)\right\rangle & = & D_{NN}\delta(\tau)\label{Lange1}\\
\left\langle\xi_F(t)\xi_F^*(t+\tau)\right\rangle & = & D_{FF}\delta(\tau)\label{Lange2},
\end{eqnarray}
\end{subequations}
where $(D_{NN},D_{FF})\in\mathbb{R}_+^2$ are respectively the diffusion coefficients of $N$ and $F$. Moreover we will assume no correlation between the two processes \cite{Dutra99, Rosencher02}
\begin{equation}
\left\langle\xi_N(t)\xi_F^*(t+\tau)\right\rangle=\left\langle\xi_F(t)\xi_N^*(t+\tau)\right\rangle=0.\label{Lange3}
\end{equation}
The identification of the second order term ($\lambda^2$) gives
\begin{EqSystem}
\frac{d \delta \Delta N^{(2)}}{d t}=-\frac{\eta}{\tau}\delta \Delta N^{(2)}-\frac{2}{\tau_{ph}}\delta F^{(2)}-2\kappa\delta \Delta N^{(1)}(t)\,\delta F^{(1)}(t)\\
\frac{d \delta F^{(2)}}{d t}=\frac{\eta-1}{2\tau}\delta \Delta N^{(2)}+\kappa\delta\Delta N^{(1)}(t)\,\delta F^{(1)}(t),
\end{EqSystem}
which shows that the nonlinear coupling between $\delta\Delta N^{(1)}(t)$ and $\delta F^{(1)}(t)$, usually neglected, gives rise to the second order of the fluctuations of $\delta \Delta N(t)$ and $\delta \Delta F(t)$. This effect is weighted by $\kappa$ which is inversely proportional to the laser mode volume. For highly confining low mode volume WGM microcavities this effect is enhanced.

\subsection{Fluctuation calculation}

\noindent By taking the Fourier transform of (\ref{Per1a}) and (\ref{Per1b}) we obtain
\begin{EqSystem}
\widehat{\delta \Delta N}^{(1)}(\omega)=\frac{1}{D(\omega)}\left[-\frac{2}{\tau_{ph}}\hat{\xi}_F(\omega)+j\omega\hat{\xi}_N(\omega)\right]\label{sol_Per1a}\\
\widehat{\delta F}^{(1)}(\omega)=\frac{1}{D(\omega)}\left[\left(\frac{\eta}{\tau}+j\omega\right)\hat{\xi}_F(\omega)+\left(\frac{\eta-1}{2\tau}\right)\hat{\xi}_N(\omega)\right]\label{sol_Per1b}
\end{EqSystem}
where we have defined
\begin{equation}
D(\omega)=\frac{\eta-1}{\tau\tau_{ph}}-\omega^2+j\omega\frac{\eta}{\tau}.
\end{equation}
We use the same method to obtain the second order terms
\begin{EqSystem}
\widehat{\delta \Delta N}^{(2)}(\omega)=-2\kappa\frac{\frac{1}{\tau_{ph}}+j\omega}{D(\omega)}\left(\widehat{\delta F}^{(1)}(\omega)\ast\widehat{\delta \Delta N}^{(1)}(\omega)\right)=K_N\left(\widehat{\delta F}^{(1)}(\omega)\ast\widehat{\delta \Delta N}^{(1)}(\omega)\right)\label{sol_Per2a}\\
\widehat{\delta F}^{(2)}(\omega)=\kappa\frac{\frac{1}{\tau}+j\omega}{D(\omega)}\left(\widehat{\delta F}^{(1)}(\omega)\ast\widehat{\delta \Delta N}^{(1)}(\omega)\right)=K_F\left(\widehat{\delta F}^{(1)}(\omega)\ast\widehat{\delta \Delta N}^{(1)}(\omega)\right)\label{sol_Per2b},
\end{EqSystem}
where $\ast$ represents the convolution product.

\subsection{RIN expression}

\noindent The RIN is defined as the normalized Fourier transform of the photon number fluctuation autocorrelation function
\begin{equation}
RIN(\omega)=\frac{\mathrm{FT}\left[\left\langle\delta F(t)\delta F^*(t+\tau)\right\rangle\right]}{\overline{F}^2}.
\end{equation}
By setting $\lambda=1$ in Eqs (\ref{pertu2}) we have to calculate four autocorrelation function Fourier transforms. Among them, using the Wiener-Khintchine theorem \cite{Mandel95} we first evaluate 
\begin{subequations}
\begin{eqnarray}
\mathrm{FT}\left[\left\langle\delta F^{(1)}(t)\left(\delta F^{(2)}(t+\tau)\right)^*\right\rangle\right] & = & \left\langle\widehat{\delta F}^{(1)}(\omega)\left(\widehat{\delta F}^{(2)}(\omega)\right)^*\right\rangle\\
            		& = & K_F^*(\omega)\left\langle\widehat{\delta F}^{(1)}(\omega)\left(\widehat{\delta F}^{(1)}(\omega)\ast\widehat{\delta \Delta N}^{(1)}(\omega)\right)^*\right\rangle.
\end{eqnarray}
\end{subequations}
Noting that
\begin{equation}
\widehat{\delta F}^{(1)}(\omega)\ast\widehat{\delta \Delta N}^{(1)}(\omega)=\mathrm{FT}\left[\delta F^{(1)}(t)\delta\Delta N^{(1)}(t)\right]
\end{equation}
and using one more time the Wiener-Khintchine theorem we obtain
\begin{equation}
\mathrm{FT}\left[\left\langle\delta F^{(1)}(t)\left(\delta F^{(2)}(t+\tau)\right)^*\right\rangle\right]= K_F^*(\omega)\mathrm{FT}\left[\left\langle\delta F^{(1)}(t)\left(\delta F^{(1)}(t+\tau)\delta\Delta N^{(1)}(t+\tau)\right)^*\right\rangle\right].
\end{equation}
Assuming that $\delta F^{(1)}(t)$ ans $\delta\Delta N^{(1)}(t)$ are Gaussian random processes, we can use the Wick-Isserlis theorem \cite{Mandel95} which gives 
\begin{equation}
\left\langle\delta F^{(1)}(t)\left(\delta F^{(1)}(t+\tau)\delta\Delta N^{(1)}(t+\tau)\right)^*\right\rangle=0,
\end{equation}
leading to the following result
\begin{equation}
\mathrm{FT}\left[\left\langle\delta F^{(1)}(t)\left(\delta F^{(2)}(t+\tau)\right)^*\right\rangle\right]=0.
\end{equation}
Using the same approach we can show that
\begin{equation}
\mathrm{FT}\left[\left\langle\delta F^{(2)}(t)\left(\delta F^{(1)}(t+\tau)\right)^*\right\rangle\right]=0.
\end{equation}
Finally we obtain 
\begin{subequations}
\begin{eqnarray}
RIN(\omega) & =& \frac{\mathrm{FT}\left[\left\langle\delta F^{(1)}(t)\left(\delta F^{(1)}(t+\tau)\right)^*\right\rangle\right]}{\overline{F}^2}+\frac{\mathrm{FT}\left[\left\langle\delta F^{(2)}(t)\left(\delta F^{(2)}(t+\tau)\right)^*\right\rangle\right]}{\overline{F}^2}\\
            & = & RIN^{(1)}(\omega)+RIN^{(2)}(\omega).\label{RIN_tot}
\end{eqnarray}
\end{subequations}
$RIN^{(2)}(\omega)$ consists of terms in $\lambda^4$. To be exhaustive, we should have developed $\delta\Delta N(t)$ and $\delta F(t)$ up to the third order in Eqs (\ref{pertu1}) and (\ref{pertu2}). Nevertheless, as we will see later, third order terms mainly contribute to a third harmonic spiking resonance which we deliberately do not consider in this work.

\subsection{First order RIN}

\noindent To obtain the expression of the RIN at first order we apply the Wiener-Khintchine theorem
\begin{equation}
RIN^{(1)}(\omega)=\frac{\mathrm{FT}\left[\left\langle\delta F^{(1)}(t)\left(\delta F^{(1)}(t+\tau)\right)^*\right\rangle\right]}{\overline{F}^2}=\frac{\left\langle\left|\widehat{\delta F}^{(1)}(\omega)\right|^2\right\rangle}{\overline{F}^2}
\end{equation}
using Eq. (\ref{sol_Per2b}) this gives
\begin{equation}\label{RIN1}
RIN^{(1)}(\omega)=\frac{1}{\overline{F}^2\left|D(\omega)\right|^2}\left[\left(\frac{\eta-1}{2\tau}\right)^2D_{FF}+\left[\left(\frac{\eta}{\tau}\right)^2+\omega^2\right]D_{NN}\right],
\end{equation}
where we used Eqs (\ref{Lange1}), (\ref{Lange2}) and (\ref{Lange3}) and the generalized Wiener-Khintchine theorem \cite{Mandel95} to write $\left\langle\left|\hat{\xi}_F(\omega)\right|^2\right\rangle=D_{FF}$, $\left\langle\left|\hat{\xi}_N(\omega)\right|^2\right\rangle=D_{NN}$ and
\begin{equation}
\left\langle\hat{\xi}_F(\omega)\hat{\xi}^*_N(\omega)\right\rangle=\left\langle\hat{\xi}_N(\omega)\hat{\xi}^*_F(\omega)\right\rangle=0.
\end{equation}
Equation (\ref{RIN1}) is the usual expression for the RIN of a class-B laser above threshold \cite{Baili08, Baili09}.

\subsection{Second order RIN}

\noindent Following the same method we obtain the second order RIN 
\begin{equation}
RIN^{(2)}(\omega)=\frac{\mathrm{FT}\left[\left\langle\delta F^{(2)}(t)\left(\delta F^{(2)}(t+\tau)\right)^*\right\rangle\right]}{\overline{F}^2}=\frac{\left\langle\left|\widehat{\delta F}^{(2)}(\omega)\right|^2\right\rangle}{\overline{F}^2},
\end{equation}
using Eq. (\ref{sol_Per2b}) this gives:
\begin{equation}
RIN^{(2)}(\omega)=\frac{\left|K_F(\omega)\right|^2}{\overline{F}^2}\left\langle \left|\widehat{\delta F}^{(1)}(\omega)\ast\widehat{\delta \Delta N}^{(1)}(\omega)\right|^2\right\rangle,
\end{equation}
which cannot be evaluated straightforwardly because it involves correlations of more complex products than thus given in Eqs (\ref{Lange1}), (\ref{Lange2}) and (\ref{Lange3}). We now introduce the function $A_2(t)=\delta F^{(1)}(t)\delta N^{(1)}(t)$ and thus
\begin{equation}\label{RIN2_def}
RIN^{(2)}(\omega)=\frac{\left|K_F(\omega)\right|^2}{\overline{F}^2}\left\langle\left|\hat{A_2}(\omega)\right|^2\right\rangle=\frac{\left|K_F(\omega)\right|^2}{\overline{F}^2}\mathrm{FT}\left[\left\langle A_2(t)A^*_2(t+\tau)\right\rangle\right].
\end{equation}
where 
\begin{Equation}
\left\langle A_2(t)A^*_2(t+\tau)\right\rangle=\left\langle\delta F^{(1)}(t)\delta N^{(1)}(t)\left(\delta F^{(1)}(t+\tau)\delta N^{(1)}(t+\tau)\right)^*\right\rangle
\end{Equation}
which can be calculated using the Wick-Isserlis theorem
\begin{Equation}
\begin{split}
\left\langle A_2(t)A^*_2(t+\tau)\right\rangle=\left\langle\delta F^{(1)}(t)\delta N^{(1)}(t)\right\rangle\left\langle\left(\delta F^{(1)}(t+\tau)\delta N^{(1)}(t+\tau)\right)^*\right\rangle+\\
\left\langle\delta F^{(1)}(t)\left(\delta F^{(1)}(t+\tau)\right)^*\right\rangle\left\langle\delta N^{(1)}(t)\left(\delta N^{(1)}(t+\tau)\right)^*\right\rangle+\\
\left\langle\delta F^{(1)}(t)\left(\delta N^{(1)}(t+\tau)\right)^*\right\rangle\left\langle\delta N^{(1)}(t)\left(\delta F^{(1)}(t+\tau)\right)^*\right\rangle\label{Isserlis}
\end{split}
\end{Equation}
The first term on the right of Eq. (\ref{Isserlis}) is constant and thus we will note its Fourier transform $C\delta(\omega)$ with $C\in\mathbb{R}_+$. We now calculate the Fourier transform of $\left\langle A_2(t)A^*_2(t+\tau)\right\rangle$
\begin{Equation}
\begin{split}
\left\langle\left|\hat{A_2}(\omega)\right|^2\right\rangle=C\delta(\omega)\\
+\left(\mathrm{FT}\left[\left\langle\delta F^{(1)}(t)\left(\delta F^{(1)}(t+\tau)\right)^*\right\rangle\right]\right)\ast\left(\mathrm{FT}\left[\left\langle\delta N^{(1)}(t)\left(\delta N^{(1)}(t+\tau)\right)^*\right\rangle\right]\right)\\
+\left(\mathrm{FT}\left[\left\langle\delta F^{(1)}(t)\left(\delta N^{(1)}(t+\tau)\right)^*\right\rangle\right]\right)\ast\left(\mathrm{FT}\left[\left\langle\delta N^{(1)}(t)\left(\delta F^{(1)}(t+\tau)\right)^*\right\rangle\right]\right)
\end{split}
\end{Equation}
which can also be written
\begin{Equation}
\begin{split}
\left\langle\left|\hat{A_2}(\omega)\right|^2\right\rangle=C\delta(\omega)+\left\langle\left|\widehat{\delta F}^{(1)}(\omega)\right|^2\right\rangle\ast\left\langle\left|\widehat{\delta N}^{(1)}(\omega)\right|^2\right\rangle+\\
\left\langle\widehat{\delta F}^{(1)}(\omega)\left(\widehat{\delta N}^{(1)}(\omega)\right)^*\right\rangle\ast\left\langle\widehat{\delta N}^{(1)}(\omega)\left(\widehat{\delta F}^{(1)}(\omega)\right)^*\right\rangle.
\end{split}
\end{Equation}
Using the same rules as those used to established Eq. (\ref{RIN1}) we have
\begin{Equation}\label{A2}
\begin{split}
\left\langle\left|\hat{A_2}(\omega)\right|^2\right\rangle =C\delta(\omega)+U_1(\omega)\left[\left(\frac{\eta-1}{\tau\tau_{ph}}\right)^2D_{NN}D_{FF}+2\left(\frac{\eta}{\tau}\right)^2\left(\frac{2}{\tau_{ph}}\right)^2D_{FF}^2\right]+\\
U_2(\omega)\left[\left(\frac{\eta-1}{2\tau}\right)D_{NN}+\left(\frac{2}{\tau_{ph}}\right)D_{FF}\right]^2+\\
U_3(\omega)\left[\left(\frac{\eta-1}{2\tau}\right)^2D_{NN}^2+\left(\frac{2}{\tau_{ph}}\right)^2D_{FF}^2+\left(\frac{\eta}{\tau}\right)^2D_{NN}D_{FF}\right]+U_4(\omega)D_{NN}D_{FF},
\end{split}
\end{Equation}
where functions $U_i(\omega)$ with $i\in\{1,2,3,4\}$ are defined by
\begin{subequations}\label{Def_U}
\begin{eqnarray}
U_1(\omega) & = & \left(\left|D(\omega)\right|^{-2}\right)\ast\left(\left|D(\omega)\right|^{-2}\right)\\
U_2(\omega) & = & \left(\omega\left|D(\omega)\right|^{-2}\right)\ast\left(\omega\left|D(\omega)\right|^{-2}\right)\\
U_3(\omega) & = & \left(\left|D(\omega)\right|^{-2}\right)\ast\left(\omega^2\left|D(\omega)\right|^{-2}\right)\\
U_4(\omega) & = & \left(\omega^2\left|D(\omega)\right|^{-2}\right)\ast\left(\omega^2\left|D(\omega)\right|^{-2}\right).																					 
\end{eqnarray}
\end{subequations}
In conclusion, using Eqs (\ref{sol_Per2b}), (\ref{RIN2_def}), (\ref{A2}) and (\ref{Def_U}) the second order RIN can be evaluated which finally gives the total RIN including this contribution in Eq. (\ref{RIN_tot}). Note that in the experiments, the DC contribution giving $C\delta(\omega)$ will be filtered out and will not be taken into account in the simulations.

\section{Experimental method}

\subsection{Experimental setup}

\noindent The experimental part is devoted to the consistency checking of our model. In this purpose we used WGM low mode volume microcavities. The WGM lasers studied in this paper consist of ZBLALiP-fluoride glass microspheres doped with erbium ions. The microspheres are made by melting glass powders in a plasma torch, details on the manufacturing process can be found in Ref. \cite{Rasoloniaina2014}. For the present experiments, we used a doping rate of $0.1\mathrm{mol}\%$ and microspheres with diameters in the range $85-90~\mu\mathrm{m}$. The refractive index of ZBLALiP is $n_0=1.49$. The laser operation is obtained around $1.56~\mu\mathrm{m}$ using the $^4I_{13/2}\rightarrow^4I_{15/2}$ transition of erbium. In our case, due to the melting process the excited state lifetime of this transition can be considered such as $T_1\approx 10~\mathrm{ms}$ \cite{Huet16} and the emission cross section will be taken equal to $3\times10^{-25}~\mathrm{m}^2$ \cite{Shea07}.
\begin{figure}[ht!]
\centering\includegraphics[width=13cm]{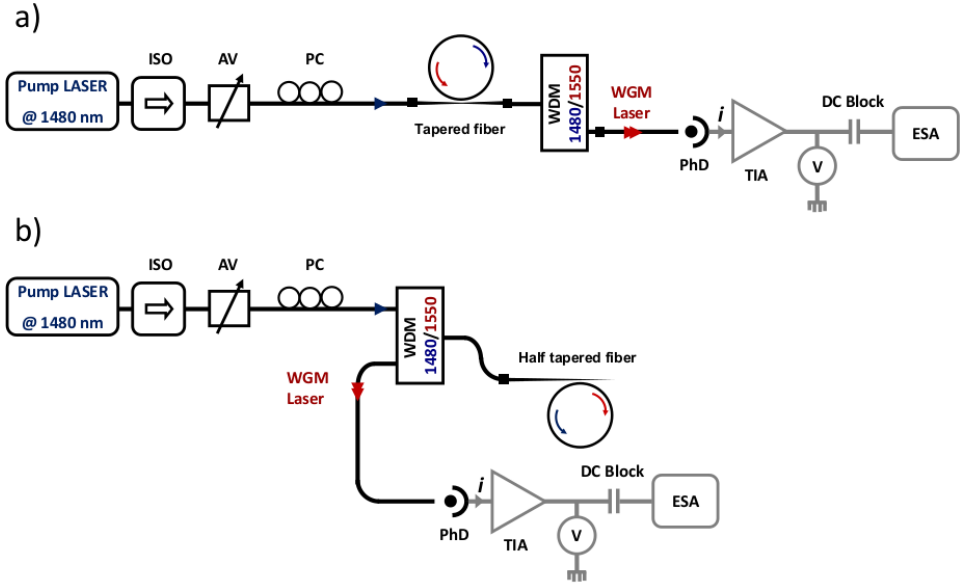}
\caption{Experimental setups. a) tapered fiber setup, b) half tapered fiber setup. ISO : optical isolator, AV : variable attenuator, PC : polarization controller, WDM : wavelength multiplexer, PhD : InGaAs photodiode, TIA : transimpedance amplifier, V : voltmeter, ESA : electrical spectrum amplifier. $i$ is the generated photo-current at the output of the photodiode.}\label{setup}
\end{figure}
The ions are resonantly pumped using WGM around $1.48~\mu\mathrm{m}$. Figure \ref{setup} is a sketched of the two setups used in our experiments. In the first setup shown in Fig. \ref{setup}.a), the pump is injected and the laser signal is extracted by a tapered fiber whose diameter is reduced to about $1~\mu\mathrm{m}$. In the second setup the pump is inserted using an half tapered fiber with a diameter reduced to less than $2~\mu\mathrm{m}$. The laser signal is collected in the counter-propagating mode using the same fiber as sketched in Fig. \ref{setup}.b). We have tested both single-mode or multi-mode pumping regime using laser diodes emitting up to $100~\mathrm{mW}$ around $1480~\mathrm{nm}$. The single-mode pumping is achieved thanks to a narrow linewidth ($<10~\mathrm{kHz}$) single-mode laser diode whereas for multi-mode pumping we use an usual broadband Fabry-Perot laser diode. The laser emission is measured using an InGaAs photodiode with a responsivity of $1~\mathrm{A/W}$ at $1550~\mathrm{nm}$ and a bandwidth of $1.2~\mathrm{GHz}$. The electrical signal is amplified using a transimpedance amplifier (TIA) with a maximal gain of $10^6~\mathrm{V/A}$ and a bandwidth of $1.2~\mathrm{MHz}$. The DC component giving the laser emitting power measured with a voltmeter (V) and filtered out using a DC block (cut-off frequency of $1~\mathrm{Hz}$).

\subsection{RIN measurement method}

\noindent From an experimental point of view, the ESA enables us to measure the power spectral density (PSD) $N_{tot}$ of the photo-current $i$ after amplification which is characterized by a transfer function $H(\omega)$. This is related to the thermal noise $N_{th}$, shot-noise $N_{sn}$ and RIN $N_{RIN}$ PSD by:
\begin{Equation}
N_{tot}=N_{th}+N_{sn}+N_{RIN}.
\end{Equation}
The experimental thermal noise PSD $N_{th}$ is measured with no signal at the photodiode ($i=0$). The shot noise PSD reads
\begin{Equation}
N_{sn}=2\left|H(\omega)\right|^2Riq
\end{Equation}
where $q$ is the elementary charge and $R$ the load resistance. This value can also be experimentally obtained using a reference source without RIN giving the same photo-current $i$ at the photodiode output. By noticing that
\begin{Equation}
N_{RIN}=R\left|H(\omega)\right|^2i^2\times RIN,
\end{Equation}
we obtain the experimental value of the RIN by
\begin{Equation}
RIN=\frac{2q}{i}\left(\frac{N_{tot}-N_{th}}{N_{sn}}-1\right),
\end{Equation}
where we deduce the photo-current $i$ from the voltage $V$ obtain thanks to the voltmeter V by $i=V/G$ where $G$ is the transimpedance gain.

\section{Results}

\begin{figure*}[ht!]
\centerline{\subfigure[Half tapered fiber]{\includegraphics[width=8.5cm]{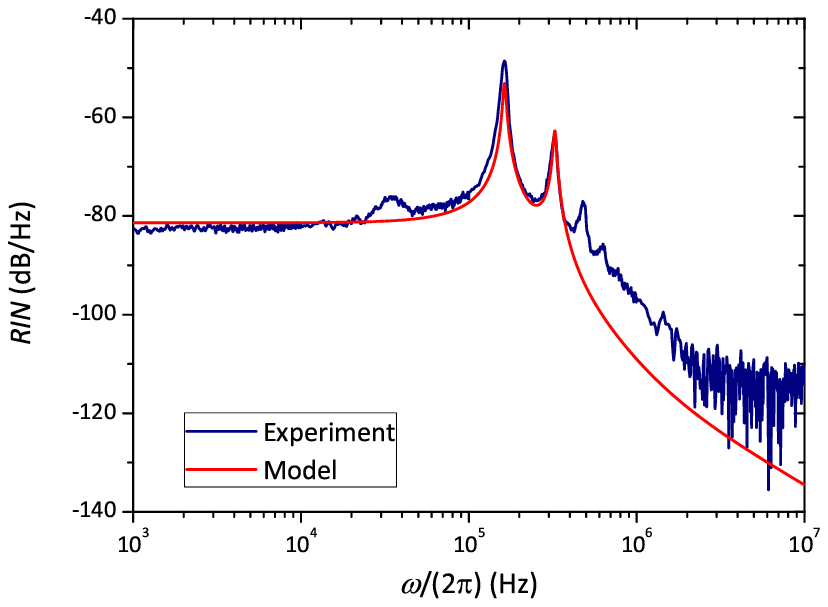}
\label{broad_a}} \hfil \subfigure[Tapered fiber]{\includegraphics[width=8.5cm]{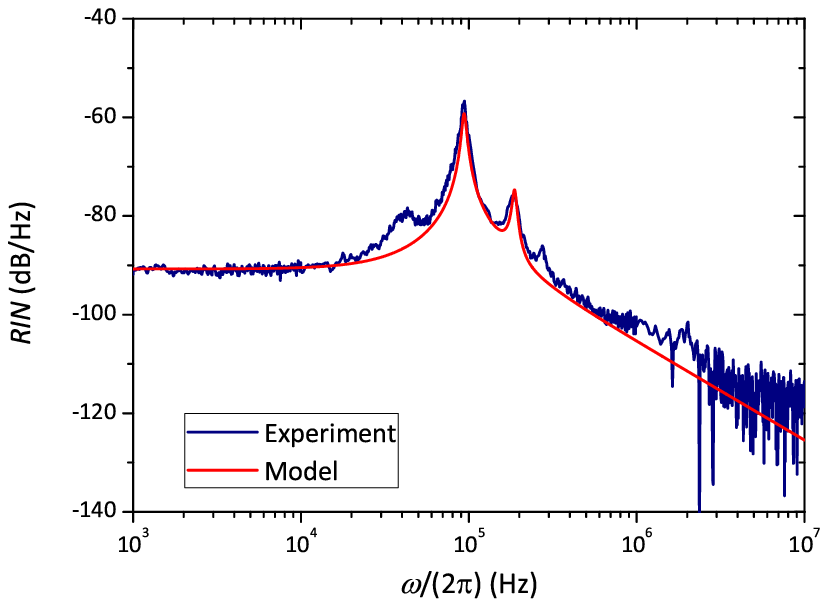}
\label{broad_b}}} \caption{RIN measurement for a broadband pumping. a) Half tapered fiber configuration and b) tapered fiber configuration. The physical and fit parameters are given in Table \ref{Tab}.}\label{broad}
\end{figure*}

\begin{figure}[ht!]
\centering\includegraphics[width=10cm]{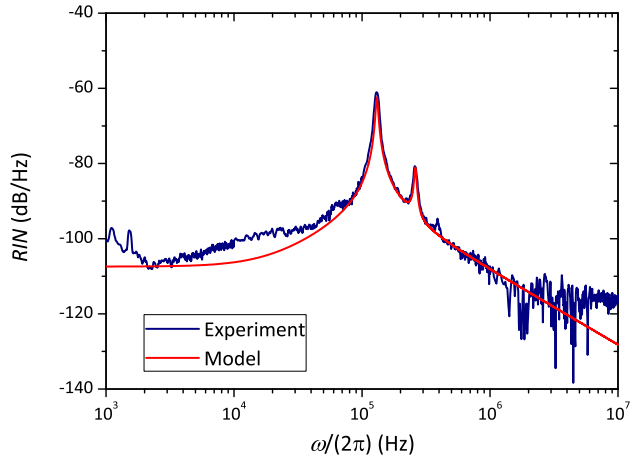}
\caption{RIN measurement in the case of a narrow linewidth pump laser and a tapered fiber configuration. The physical and fit parameters are given in Table \ref{Tab}.}\label{narrow_pump}
\end{figure}

\noindent We used the same methodology for all the results presented in this section. We compare several experimental configurations and different microspheres with comparable diameters. The RIN spectra are all recorded for the maximal power laser obtained with a single-mode operation. This is checked using a spectral analyser (confocal cavity) with a spectral resolution of $100~\mathrm{MHz}$ (not shown on Fig. \ref{setup}). Note that the single-mode operation can also be checked by making sure that only one fundamental relaxation oscillation frequency is present in the RIN spectrum. In Fig. \ref{broad}, RIN measurements are reported in the case of broadband pump laser a) in the half tapered fiber configuration and b) in the tapered fiber configuration. The RIN spectra show resonances (relaxation peak) which are the signature of a class-B laser operation. Harmonics of this relaxation oscillation frequency are also clearly visible on both spectra. Up to three harmonics ($326$, $488$, $640~\mathrm{kHz}$) of the relaxation oscillation frequency ($163~\mathrm{kHz}$) are present on Fig. \ref{broad_a}. Note that on both spectra a broad weak peak at $34~\mathrm{kHz}$ on Fig. \ref{broad_a} and $41~\mathrm{kHz}$ on Fig. \ref{broad_b} can be attributed to a second laser mode near threshold which has no influence on our analysis. The experimental data are manually fitted using as free parameters : i) the pumping rate $r$ and the photon lifetime $\tau_{ph}$ which give the position, the width and the amplitude of the fundamental resonance ii) the values of $D_{NN}$ and $D_{FF}$ influencing the global noise level and iii) the value of $\kappa$ which unambiguously sets the ratio between the fundamental and first harmonic peaks. Our second order development is in good agreement for the RIN levels both at the spiking frequency and its first harmonic. To obtain a third peak in the simulation we would have developed $\delta N(t)$ an $\delta F(t)$ at the third order. In this case, the products of first and third order terms would give contributions at the first harmonic with amplitude similar to the third peak which should be neglected. Numerical values of the fitting parameters are given in Table \ref{Tab}. The external Q-factor $Q_e$ can be obtained from the operation wavelength $\lambda_0$ and the photon lifetime by $Q_e=2\pi c\tau_{ph}/\lambda_0$ and is also given in Table \ref{Tab}.
These values are in the range of $10^5-10^9$ which can be achieved with our experimental setup \cite{Rasoloniaina12, Huet16}. WGM are labelled using three quantum numbers $(n,\ell,m)$. $n$ is the radial order of the mode, $\ell$ is the interference order and the azimuthal number $\ell-\left|m\right|+1$ is such as it gives the number of maxima of the field in the polar direction. The volume of the WGM is deduced from $\kappa$ (see Table \ref{Tab}) using Eq. (\ref{cappa}). By comparing this value to the theoretical model \cite{Collot93} 
\begin{equation}
V_m=\frac{\iiint_{\mathbb{R}^3}w(\mathrm{\textbf{r}})\mathrm{d^3\textbf{r}}}{\mathrm{max}\left[w(\mathrm{\textbf{r}})\right]}
\end{equation}
where $w(\mathrm{\textbf{r}})$ is the electromagnetic energy density, we can deduce the quantum numbers associated to the laser WGM. For the microsphere corresponding to Fig. \ref{broad_a}, two sets of quantum numbers $(n,\ell-\left|m\right|)$ are found: $(1,5)$ or $(2,1)$, whereas for Fig. \ref{broad_b} we obtain the pair $(1,2)$. These quantum numbers are typical values of what can be characterized with our experimental setup \cite{Rasoloniaina2015}. Since the tapered fiber configuration seems to produce less noise at low frequencies, we tested it with a narrow linewidth pump laser. The results are given in Fig. \ref{narrow_pump}.
The fit gives an external Q-factor $Q_e=5.7\times 10^7$ and a WGM with $n=2$ and $\ell-\left|m\right|=1$. Again, these values are totally consistent with what is usually observed with this kind of WGM microcavity. Finally, in this pumping configuration, the noise at low frequencies is reduced by one order of magnitude in comparison with the broadband pumping. This effect could come from the fact that, for a narrow pump laser, less non-efficient WGM pump modes are excited resulting in a lower excess noise. This assumption could also been done for the reduction of the low frequency noise when using a tapered fiber which allows a more transverse-mode selective pumping. As a remark, the RIN measured in WGM lasers at low frequency is 3 orders of magnitude higher than that of fiber lasers \cite{Loh98}; at the spiking frequency, the RIN can be even 6 orders of magnitude greater.
\begin{table}[h!]
  \caption{Physical and fit parameters for the WGM laser shown in the three previous figures. $a$~:~radius of the sphere, $P_L$ : WGM laser power. We also give the deduced parameters $Q_e$ and $V_m$. The mode volume is calculated in TE since the theoretical resonance wavelength values match well with experimental data in this polarization. Error bars on the fitting parameters are arbitrarily set using the following method. For $r$, variations of about $1~\%$ gives a change of $1~\mathrm{kHz}$. $10~\%$ error on $\tau_{ph}$ induces variations of $1.5~\mathrm{dB}$ on the relaxation peak. Variations up to $30~\%$ gives only $1~\mathrm{dB}$ variation in the low frequency white noise. $D_{FF}$ influences the value of the amplitude of the relaxation peak, a $20~\%$ variation induces $1~\mathrm{dB}$ variation of the peak. Finally, a change in $\kappa$ of $10~\%$ gives a variation of $1~\mathrm{dB}$ in the relative position of the fundamental and harmonic resonances.} \label{Tab}
  \begin{center}
	\renewcommand{\arraystretch}{1}
    \begin{tabular}{c c c c}
    \hline
    & Fig. \ref{broad_a} & Fig. \ref{broad_b} & Fig. \ref{narrow_pump}\\
		\hline
		$2a~(\mu\mathrm{m})$ & $85$ & $90$ & $90$\\
		$P_L~(\mathrm{nW})$ & $57.0$ & $105$ & $129$\\
		$\lambda_0~(\mathrm{nm})$ & $1563.6$ & $1565.1$ & $1565.7$\\
		\hline
		$r$ & $9.5\pm 0.1$ & $3.8\pm 0.1$ & $6.5\pm 0.1$\\
		$\tau_{ph}~(\mathrm{ns})$ & $40\pm 4$ & $(1.0\pm0.1)\times 10^2$ & $47\pm 5$\\
		$D_{NN}~(\mathrm{s}^{-1})$ & $(2.0\pm 0.7)\times 10^{19}$ & $(2.5\pm 0.8)\times 10^{17}$ & $(1.5\pm 0.5)\times 10^{16}$\\
		$D_{FF}~(\mathrm{s}^{-1})$ & $(1.5\pm 0.3)\times 10^{14}$ & $(8.5\pm 1.7)\times 10^{14}$ & $(3.0\pm0.6)\times 10^{14}$\\
		$\kappa~(\mathrm{s}^{-1})$ & $(2.0\pm0.2)\times 10^{-2}$ & $(2.0\pm0.2)\times 10^{-2}$ & $(2.2\pm0.2)\times 10^{-2}$\\
    \hline
		$Q_e$ & $(4.8\pm 0.5)\times 10^7$ & $(1.2\pm 0.1)\times 10^8$ & $(5.7\pm 0.6)\times 10^7$\\
		$V_m~(\mu\mathrm{m}^{3})$ & $(3.0\pm 0.6)\times 10^3$ & $(3.0\pm 0.6)\times 10^3$ & $(2.7\pm 0.5)\times 10^3$\\
		\hline
    \end{tabular}
  \end{center}
\end{table}

\section{Conclusion}

\noindent We developed a RIN model for class-B lasers including nonlinear coupling of population inversion and photon number fluctuations. This model is well suited for the description of low mode volume WGM lasers for which this coupling is not negligible. The model has been checked by characterizing noise properties of erbium-doped glass microspheres under several pumping configuration. We found that single-mode pumping using a tapered fiber is the most promising configuration for getting small low frequency intensity noise. The parameters obtained from the comparison between experiments and our model are consistent. In particular we have shown that the mode volume and the quantum numbers of the WGM laser could be deduced from the RIN spectrum analysis. This can be useful in the case of strongly multimode WGM microcavities for which dedicated experiments must be implemented to obtain a full spatial mode characterization. Furthermore, the fitting process could be greatly improved using an automatic routine and developed to higher orders. Our approach could be extended to WGM laser frequency noise and exploited in the analysis of the whole noise properties of micro-lasers used in sensor applications.

\section*{Acknowledgments}
\noindent J.-B. Ceppe acknowledges support from the Centre National d'\'{E}tudes Spatiales (CNES) and R\'{e}gion Bretagne (ARED). Y. Dumeige is member of the Institut Universitaire de France. The authors acknowledge fruitful discussions with T. Chartier and S. Trebaol.

\end{document}